\documentclass[usenatbib,
               useAMS]{mnras}

\usepackage{placeins,graphicx,amsmath,amssymb}

\title[The Antesonic Condition I]{The Antesonic Condition for the Explosion of Core-Collapse Supernovae I: Spherically Symmetric Polytropic Models: Stability \& Wind Emergence}
\author[M. J. Raives et al.]{Matthias J. Raives$^{1}$,
                             Sean M. Couch$^{2,3,4,5}$,\newauthor{}
                             Johnny P. Greco$^{6}$,
                             Ond\v{r}ej Pejcha$^{7}$,
                             Todd A. Thompson$^{1,8}$,
                             \\
                             ${}^{1}$Department of Astronomy, The Ohio State University, Columbus, OH 43210, USA\\
                             ${}^{2}$Department of Physics and Astronomy, Michigan State University, East Lansing, Mi 48824, USA\\
                             ${}^{3}$Department of Computational Mathematics, Science, and Engineering, Michigan State University, East Lansing, Mi 48824, USA\\
                             ${}^{4}$National Superconducting Cyclotron Laboratory, Michigan State University, East Lansing, Mi 48824, USA\\
                             ${}^{5}$Joint Institute for Nuclear Astrophysics, Michigan State University, East Lansing, Mi 48824, USA\\
                             ${}^{6}$Department of Astrophysical Sciences, Princeton University, Princeton, NJ 08540, USA.\\
                             ${}^{7}$Institute of Theoretical Physics,Faculty of Mathematics and Physics, Charles University, Prague, Czech Republic\\
                             ${}^{8}$Center for Cosmology and Astroparticle Physics, The Ohio State University, Columbus, OH 43210, USA
                             }
\begin{document}

    \maketitle

    \begin{abstract}
        Shock revival in core-collapse supernovae (CCSNe) may be due to the neutrino mechanism. While it is known that in a neutrino-powered CCSN, explosion begins when the neutrino luminosity of the proto-neutron star exceeds a critical value, the physics of this condition in time-dependent, multidimensional simulations are not fully understood.  \citet{Pejcha2012} found that an `antesonic condition' exists for time-steady spherically symmetric models, potentially giving a physical explanation for the critical curve observed in simulations.
        In this paper, we extend that analysis to time-dependent, spherically symmetric polytropic models. We verify the critical antesonic condition in our simulations, showing that models exceeding it drive transonic winds whereas models below it exhibit steady accretion.  In addition, we find that (1) high spatial resolution is needed for accurate determination of the antesonic ratio and shock radius at the critical curve, and that low resolution simulations systematically underpredict these quantities, making explosion more difficult at lower resolution; (2) there is an important physical connection between the critical mass accretion rate at explosion and the mass loss rate of the post-explosion wind: the two are directly proportional at criticality, implying that, at criticality, the wind kinetic power is tied directly to the accretion power; (3) the value of the post-shock adiabatic index $\Gamma$ has a large effect on the length and time scales of the post-bounce evolution of the explosion larger values of $\Gamma$ result in a longer transition from the accretion to wind phases.
    \end{abstract}
    \begin{keywords}
        accretion, accretion disks -- hydrodynamics -- shock waves -- supernovae: general
    \end{keywords}

    \section{Introduction}
        When the iron core of a massive star collapses, the collapse is halted as the core exceeds nuclear densities, driving a shockwave into the infalling progenitor.  This shockwave stalls, becoming a standing accretion shock at scales $r\sim200\:\mathrm{km}$, and the progenitor continues to accrete onto the proto-neutron star (PNS) until the shock is revived by neutrino heating, leading to explosion, or until a black hole is formed \citep[e.g.,][]{Bethe1985,Herant1994,Burrows1995,Janka1995}.  Much of supernova theory is focused on understanding the revival of the stalled shock, especially in 2D and 3D \citep{Couch2013,Dolence2015,OConnor2018}.

        The idea of a critical neutrino luminosity for supernovae was first explored in \citet{Burrows1993} (hereafter, BG93). They showed that for a spherical, time-steady accretion flow with optically-thin  neutrino heating and cooling, for a given accretion rate $\dot{M}_{\mathrm{acc}}$ there exists a critical core neutrino luminosity above which no steady-state accretion solution can be found.  This work has since been extended to find 2-dimensional (2D) and 3-dimensional (3D), as well as time-dependent critical curves and critical surfaces \citep{Murphy2017}.  Such studies have generally found that the 2D and 3D critical curves have a smaller normalization than the 1-dimensional (1D) curve; i.e., the critical luminosity for a given $\dot{M}_{\mathrm{acc}}$ is smaller in 2D and 3D than it is in 1D (but larger in 3D than in 2D) \citep{Murphy2008, Couch2013,Takiwaki2014}.  Furthermore, time-dependent studies have found radial \citep{Fernandez2012,Gabay2015} and non-radial \citep{Yamasaki2005,Yamasaki2007} instabilities such as the standing accretion shock instability \citep{Blondin2003,Foglizzo2006,Murphy2008,Fernandez2009,Fernandez2015}  and neutrino-driven convection \citep{Murphy2011,Murphy2013} that may tend to lower the critical curve, potentially facilitating explosions.

        In an effort to explain the existence of the critical neutrino luminosity of BG93, \citet{Pejcha2012} (hereafter, PT12) investigated the critical condition for explosion using a time-steady model, and with different levels of approximation for the post-shock microphysics and thermodynamics.  For the simple toy model of pressureless free fall onto a standing shockwave and an isothermal post-shock medium, they found an analytic critical condition -- the ratio of the isothermal sound speed to the escape velocity at the shock cannot exceed a critical threshold:
        \begin{equation}
            \xi^{\mathrm{iso}}_{\mathrm{crit}}\equiv\left.\frac{c_{T}^{2}}{v_{\mathrm{esc}}^{2}}\right|_{\mathrm{shock}} = \frac{3}{16}.
        \end{equation}
        PT12 call this critical condition the `antesonic condition,' because the condition is met at smaller radius than the sonic condition (in an isothermal wind, $c_{T}^{2}/v_{\mathrm{esc}}^{2}=\frac{1}{4}$) is. We refer to the ratio of the sound speed squared to the escape velocity squared as the `antesonic ratio,' which we denote as $\xi$.

        PT12 extended their analysis to both time-steady polytropic models and models with a general equation of state and neutrino heating and cooling (i.e., the BG93 problem), numerically deriving antesonic conditions for each case.  For polytropic models they numerically derive $\xi_{\mathrm{crit}}^{\mathrm{poly}}\simeq0.19\Gamma$.  In Appendix~\ref{app:antesonic}, we provide an analytic derivation which shows that:
        \begin{equation}
            \xi^{\mathrm{poly}}_{\mathrm{crit}}\equiv\left.\frac{c_{s}^{2}}{v_{\mathrm{esc}}^{2}}\right|_{\mathrm{max}} = \frac{3}{16}\Gamma.
            \label{eq:antepoly}
        \end{equation}
        For the general case with neutrino heating and cooling, PT12 found that
        \begin{equation}
            \xi^{\nu}_{\mathrm{crit}}\equiv\left.\frac{c_{s}^{2}}{v_{\mathrm{esc}}^{2}}\right|_{\mathrm{max}} \simeq 0.19.
            \label{eq:antenu}
        \end{equation}

        Here $\Gamma$ is the adiabatic index, $c_{s}$ is the adiabatic sound speed, and pressureless free-fall upstream of the shock is assumed.  In these more general cases, the critical condition is on the maximum value of the antesonic ratio, which may or may not be at the shock radius.  For models with neutrino heating and cooling, for example, the antesonic ratio reaches its maximum near the `gain' radius \citep{Pejcha2012}, where neutrino heating balances cooling in the post-shock flow.

        As shown by PT12, for the isothermal, polytropic, or general equation of state (EOS) with neutrino heating and cooling, or with any arbitrary changes to the heating and cooling physics, the physics of the antesonic condition is the same: above the critical antesonic ratio, it is impossible to simultaneously satisfy both the Rankine-Hugoniot shock-jump conditions and the spherically symmetric, time-steady Euler equations.  PT12 associate the antesonic condition with a \emph{dynamic} transition from steady spherical accretion to a transonic thermal wind, i.e., the supernova explosion.

        Although the antesonic condition of PT12 provides an explanation for the critical curve, it has not been fully examined in time-dependent or multi-dimensional simulations.  However, the antesonic condition has been used to predict outcomes for large suites of massive star progenitors \citep{Pejcha2015} in qualitative agreement with 1-dimensional time-dependent simulations tuned to produce explosions in 1987A-like progenitors \citep{Ugliano2012,Ertl2016,Sukhbold2016}.  It has also been used to characterize the results of multi-dimensional simulations \citep{Dolence2013,Couch2013b,Couch2014}, suggesting a critical value of $\xi^{\nu}_{\mathrm{crit}}\sim0.2-0.3$ in 2D and 3D.  Even so, there is not yet a study that shows how the dynamical transition from accretion to explosion actually occurs in a time-dependent system, in the context of the antesonic condition.

        In this paper, we take a step forward in understanding the nature of the critical curve in idealized, but time-dependent simulations, in order to better understand the nature of the transition from accretion to explosion that occurs when the antesonic ratio is exceeded -- how this process behaves, how the structure of our model evolves -- as a guide for full-physics simulations and as a test of our understanding of the physics in a simplified context.  We adopt an approach similar to the model problems explored in PT12.  In particular, we adopt a simplified, polytropic equation of state (EOS) in order to better understand the relevant physics in the accretion region, but we extend the results of PT12 to time-dependent simulations.

        This paper is organized as follows.  In section \S\ref{sec:Methods}, we describe our computational methodology.  In section \S\ref{sec:Results}, we present our results.  In \S\ref{sec:TAC}, we establish the applicability of the time-steady antesonic condition to our simulations -- i.e., that models that exceed $\xi_{\mathrm{crit}}^{\mathrm{poly}}$ explode and that models that do not exceed $\xi_{\mathrm{crit}}^{\mathrm{poly}}$ do not.  In \S\ref{sec:Resolution}, we show the resolution dependence of our simulations -- specifically, we show that low resolution models underpredict $\xi_{\mathrm{crit}}^{\mathrm{poly}}$ and $R_{\mathrm{sh}}$ at the critical curve.  In \S\ref{sec:TTW}, we describe the properties of the transonic wind -- in particular, how the wind mass loss rate is determined by the accretion rate at explosion -- which establishes an explicit connection between the accretion power at the shock and the energy of the explosion.  In \S\ref{sec:WDS}, we describe the emergence of a wind-driven shell as the wind sweeps up the accreting matter.  In \S\ref{sec:TDP}, we explore the effects of time-dependent perturbations to the simulations, i.e., whether a model can explode when it only temporarily exceeds the critical curve.  In \S\ref{sec:Conclusion}, we provide a brief conclusion.

        \begin{figure*}
            \centering{}
            \includegraphics[width=\linewidth]{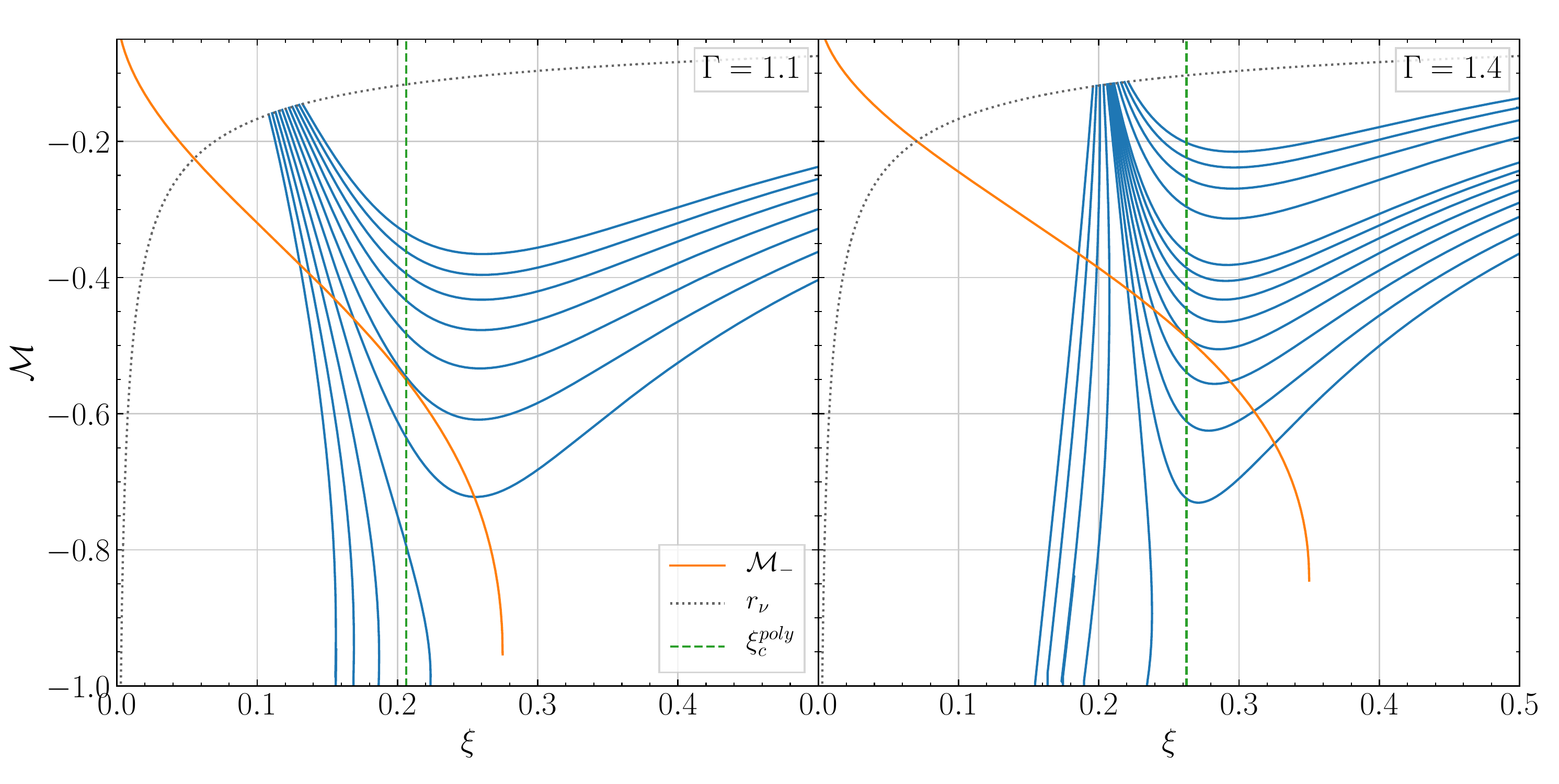}
            \caption{Solutions to the time-steady Euler equations, plotted as the Mach number $\mathcal{M}$ versus the antesonic ratio $\xi$, for $\Gamma=1.1$ (left) and $\Gamma=1.4$ (right).  The dashed gray curve shows the neutrinosphere radius $r_{\nu}$.  The blue curves correspond to configurations with different choices of $K$ at fixed $\dot{M}_{\mathrm{acc}}$.  For $\Gamma=1.1$, we show profiles of $|\delta K|$ (the fractional separation from the critical curve, see \autoref{eq:deltaK}) from 0 to $10^{-1}$, with step sizes of $10^{-2}$.  For $\Gamma=1.4$, we show $|\delta K|$ from 0 to $6\times10^{-2}$, in steps of $3\times10^{-3}$ from 0 to $1.2\times10^{-2}$, then in steps of $1.2\times10^{-2}$ up to $6\times10^{-2}$.  The orange curve shows the Mach number $\mathcal{M}_{-}$ immediately downstream of the shock (\autoref{eq:machminus}); the shock is located where the blue and orange curves intersect.  The critical $K$ for a given $\dot{M}_{\mathrm{acc}}$ is the $K$ such that the resultant velocity profile is tangent to the post-shock Mach number curve at the point of intersection.  Velocity profiles that do no intersect the $\mathcal{M}_{-}$ curve do not have an accretion shock, and thus are not considered here.  Some profiles have two points of intersection with the $\mathcal{M}_{-}$ curve; in these cases, the shock is located at the first intersection (i.e., the intersection at smaller $\xi$, and therefore $r$).}
            \label{fig:Euler}
        \end{figure*}

    \section{Methodology}\label{sec:Methods}

        Our study consists of a set of 1D, spherically symmetric hydrodynamics simulations.  The time evolution of the system is governed by the hydrodynamic equations:
        \begin{align}
            \frac{\partial \rho}{\partial t} + \vec{\nabla}\cdot(\rho\vec{v}) &= 0,\label{eq:eulerc}\\
            \frac{\partial (\rho \vec{v})}{\partial t} + \vec{\nabla}\cdot(\rho\vec{v}\otimes\vec{v}) + \vec{\nabla}P -\rho\vec{g} &= 0,\label{eq:eulerm}
        \end{align}
        where $\rho$ is the gas density, $v_{r}=\vec{v}\cdot\hat{r} = |\vec{v}|$ is the radial velocity, $P$ is the gas pressure, and $\vec{g}=-{GM}/{r^{2}}$ is the gravitational force, where $G$ is the gravitational constant, and $M=1.4\:$M$_{\odot}$ is the mass of the central PNS; we do not consider the self-gravity of the gas in our simulations.  We use a simple, polytropic EOS for the flow downstream of the shock:
        \begin{equation}
            P = K\rho^{\Gamma}
        \end{equation}
        where $\Gamma$ is the adiabatic index and $K$ is a normalization factor, where $\ln(K)$\footnote{Throughout this paper, we use ``$\ln$'' to denote the natural logarithm and ``$\log$'' to denote the base 10 logarithm.} is proportional to the entropy (for an ideal gas).  We find that, in 1D supernova simulations taking the full microphysics, heating, and cooling into account, $\Gamma$ ranges from $\Gamma\sim1.5$ near the PNS surface to $\Gamma\sim1.2$ outside the shock.  Thus, our choice of $\Gamma=1.1,1.4$ should be seen as limiting cases of the thermodynamic properties of the accretion flow. $K$ can be considered analogous to the critical neutrino luminosity $L_{\nu,\mathrm{crit}}$ considered in other studies \citep{Pejcha2012}.  However, the connection between $K$ and $L_{\nu,\mathrm{crit}}$ has not been examined in detail, and a full treatment of such is outside the scope of this paper.

        We solve the hydrodynamics equations with the \textsc{flash}\footnote{Available for download at www.flash.uchicago.edu.} code \citep{Fryxell2000}. Our fiducial simulations use the directionally unsplit hydrodynamics solver, third-order piecewise parabolic spatial reconstruction, the `hybrid' slope limiter, and the LLF Riemann solver (we discuss alternate hydro solvers in \S\ref{sec:Resolution}).  The fiducial simulations were run on a grid with inner radius $x_{\mathrm{min}}=r_{\nu}=30\:$km and outer radius $x_{\mathrm{max}}=5000\:$km, with a minimum grid spacing $\delta x \approx 0.607\:$km, obtained using a maximum of 8 AMR refinement levels.  The level of refinement decreases outward with radius in an approximately logarithmic manner. The initial shock is completely contained within the highest level of refinement.  The number of zones behind the shock (at a given resolution) is a function of the initial shock radius, and thus the EOS parameter $K$ and the mass accretion rate $\dot{M}_{\mathrm{acc}}$.  For $\Gamma=1.1$, the number of zones ranges from about 60 to 250, and for $\Gamma=1.4$, the number of zones ranges from about 60 to 4500.  The number of zones is maximized for small $\dot{M}_{\mathrm{acc}}$ and K close to the critical value.  In the case of successful explosions, the shock moves outward and into regions of lower refinement. Special care is taken when the shock encounters forced decrements in refinement to avoid under/overshooting in the interpolation and to guarantee conservation.  The effects of numerical resolution on our results are explored in more detail in \S\ref{sec:Resolution}.

        Inside the shock radius, the gas is initialized to the steady-state velocity and density profiles determined in PT12.  These profiles are defined by $K$, $\Gamma$ and the mass accretion rate $\dot{M}_{\mathrm{acc}}$.  The conditions at the shock are described by the Rankine-Hugoniot shock-jump conditions:
        \begin{align}
            \rho_{-}v_{-} &= \rho_{+}v_{+}\\
            \rho_{-}v_{-}^{2} + P_{-} &= \rho_{+}v_{+}^{2} + P_{+},
        \end{align}
        where $+$ and $-$ denote quantities just upstream and just downstream of the shock, respectively.  We use just these two conditions, neglecting the enthalpy condition, because our choice of $K$ both upstream and downstream of the shock (see below) fixes the compressibility and entropy of the post-shock medium, rendering the enthalpy condition moot.  In a self-consistent calculation with neutrino heating and cooling, the enthalpy of the immediate post-shock medium would be determined by the full set of shock-jump conditions.  Our use of a polytropic EOS with prescribed post-shock $K$ obviates the need for the third shock-jump condition and allows us to control the thermal properties of the accreting material by hand.

        Outside the shock radius, the density and velocity of the gas are set by the equations of pressureless free-fall:
        \begin{align}
            v_{+}(r) &= \sqrt{\frac{2GM}{r}},\\
            \rho_{+}(r) &= \frac{\dot{M}_{\mathrm{acc}}}{4\pi r^{2}v(r)} = \frac{\dot{M}_{\mathrm{acc}}}{4\pi r^{3/2}}\frac{1}{\sqrt{2GM}}.
        \end{align}
        For numerical reasons, instead of using $P_{+}=0$, we switch from a small but finite $K$ upstream of the shock to a large, specified, $K$ downstream of the shock.  The relevant solution to the shock-jump conditions (assuming $P_{+}=0$) is then:
        \begin{equation}
            \mathcal{M}_{-}\equiv\frac{v_{-}}{c_{s}} = \frac{1}{2}\left(\sqrt{\frac{1}{\xi}-\frac{4}{\Gamma}}-{\sqrt{\frac{1}{\xi}}}\right){},
            \label{eq:machminus}
        \end{equation}
        where all quantities are evaluated just downstream of the shock.  Here, $\xi$ is the local antesonic ratio,
        \begin{equation}
            \xi = \frac{c_{s}^{2}}{v_{\mathrm{esc}}^{2}} = \frac{K\Gamma\rho^{\Gamma-1}r}{2GM},
        \end{equation}
        as opposed to the critical antesonic ratio $\xi_{\mathrm{crit}}$, which is the \emph{maximum} value of the antesonic ratio (see \autoref{eq:antepoly}).

        We choose boundary conditions such that, at the outer boundary, the velocity and density profiles are consistent with free-fall and a constant mass accretion rate.  For other variables, zero-gradient or ``outflow'' boundary conditions are enforced.  For the inner boundary, we use a “fixed” boundary condition that is constant in time and taken from the initial conditions at the inner radius determined by the starting PT12 profile.

        We show the time-steady solutions to this problem (\autoref{eq:eulerc} and \autoref{eq:eulerm} with $\Gamma=1.1,1.4$) in \autoref{fig:Euler}.  The solid blue lines correspond to different values of $K$ in the post-shock medium, and the solid orange line is the Mach number just downstream of the shock, given in \autoref{eq:machminus}.  For a given profile, the shock exists at the intersection between the blue and orange curves.  For cases where two points of intersection exist, the shock is located at the interior one, i.e., at smaller $\xi$, and thus smaller $r$ at fixed $K$ \citep{Pejcha2012}.

        In all of our simulations, we begin by specifying $K,\dot{M}_{\mathrm{acc}}$ from a grid of values spaced linearly in $K$ and logarithmically in $\dot{M}_{\mathrm{acc}}$, in the interval $0.1$ M$_{\odot}\:$s$^{-1}\leq\dot{M}_{\mathrm{acc}}\leq1.06$ M$_{\odot}\:$s$^{-1}$.  This choice defines the initial, time-steady, density and velocity profiles. We allow the system to equilibrate for $0.1\:$s, then decrease the mass accretion rate by a factor $f_{\dot{M}}$ (which may be equal to 1) for a time $\Delta t_{\dot{M}}$ (which may be equal to the total simulation length).  The full simulation length is $2.0\:$s.

    \section{Results}\label{sec:Results}

        \subsection{The Antesonic Condition}\label{sec:TAC}
\begin{figure}
                \centering{}
                \includegraphics[width=\linewidth]{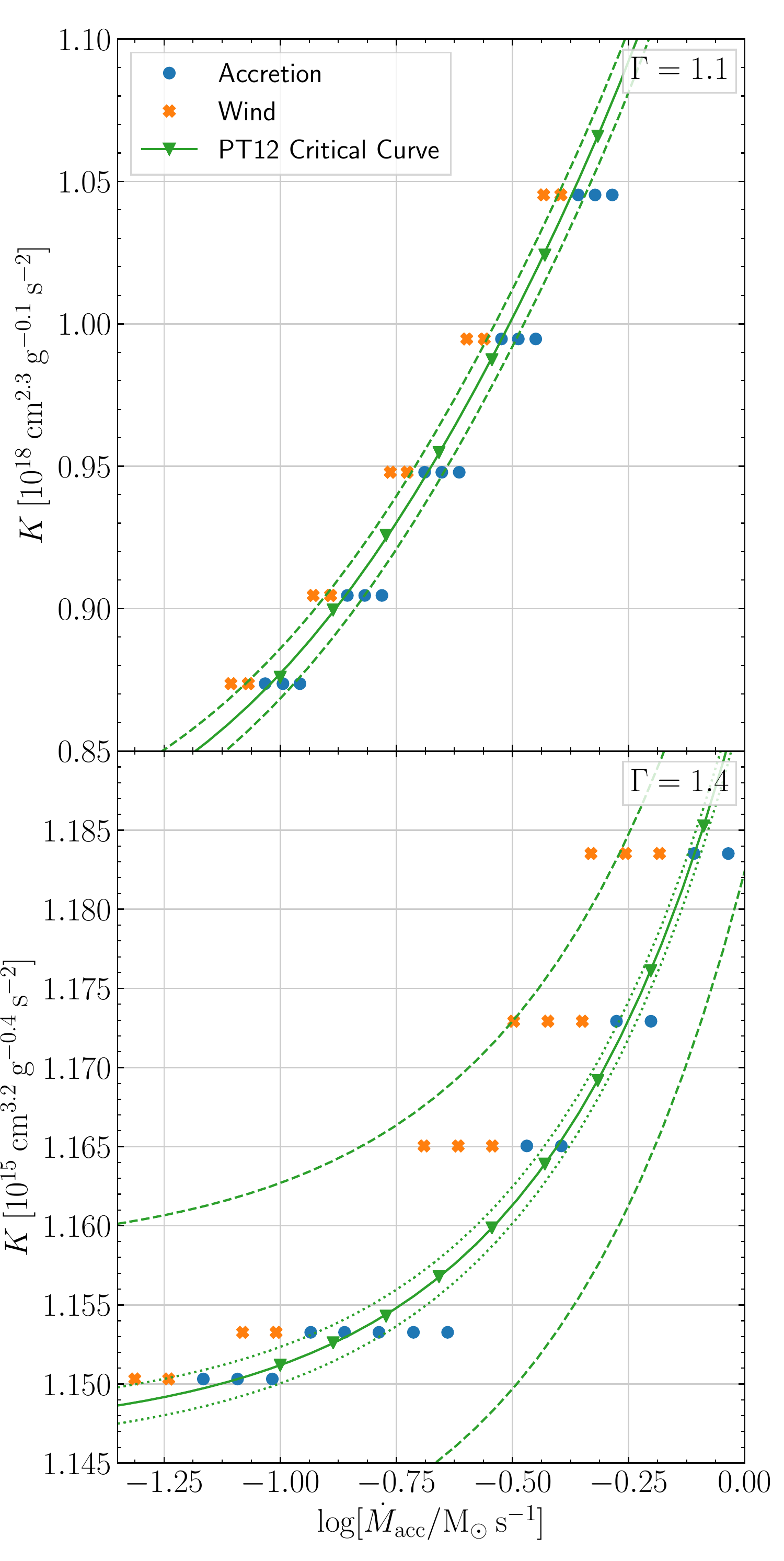}
                \caption{The critical curve for our parametrized supernova model, for $\Gamma=1.1$ (top) and $\Gamma=1.4$ (bottom), run at the fiducial resolution (with minimum grid spacing $\delta x = 0.607\:\mathrm{km}$).  Green lines show curves of constant $\delta K=0$ (solid; i.e., the critical curve), $\delta K=\pm10^{-2}$ (dashed), and $\delta K=\pm10^{-3}$ (dotted; $\Gamma=1.4$ only); green triangles show the PT12 profiles themselves.  Blue dots and orange X's represent accretion and wind solutions, respectively.  For a given accretion rate $\dot{M}_{\mathrm{acc}}$, values of $K$ above a certain critical value result in a wind solution we identify with a supernova.  We note that, for $\Gamma=1.4$, some configurations that lie above the critical curve do not explode; we attribute this to the resolution dependence of the critical curve (see \S\ref{sec:Resolution} for more details).}
                \label{fig:CriticalCurve}
            \end{figure}

            We first reproduce an approximation of the PT12 critical curve, as shown in \autoref{fig:CriticalCurve}.  We start the simulations at some stable accretion rate $\dot{M}_{\mathrm{acc},0}$, which we then decrease by constant factors $f_{\dot{M}}$ at the outer boundary.  For $(\dot{M},K)$ configurations that lie below the critical curve, the simulations maintain the initial conditions, displaying time-steady accretion solutions.  As we move the simulations to configurations that lie above the critical curve by decreasing $\dot{M}_{\mathrm{acc}}$, they undergo dynamical transformations to time-steady wind solutions.  We identify the wind solutions with successful supernova explosions.

            For the unstable configurations, we investigate the time required for explosion.  As there is no unique, well-defined way to identify the time for the onset of explosion, we consider three such definitions.  $t_{\mathrm{ante}}$ is defined to be the first time for which $\xi>\xi_{\mathrm{crit}}^{\mathrm{poly}}$ (see \autoref{eq:antepoly}); we use this time-scale as a physically motivated definition of $t=0$.  $t_{\mathrm{wind}}$ is defined as the first time for which the fluid velocity immediately downstream of the shock is positive.  $t_{400}$ is defined to be the time at which the shock reaches a fixed radius $r=400\:$km.  Though this definition is more arbitrary than the other two, it mirrors definitions used in the literature, e.g., \citet{Couch2013}.  For comparison, we also consider $t_{\mathrm{sonic}}$, the time at which the wind first achieves $\mathcal{M}=1$.

            \begin{figure*}
                \centering{}
                \includegraphics[width=\textwidth]{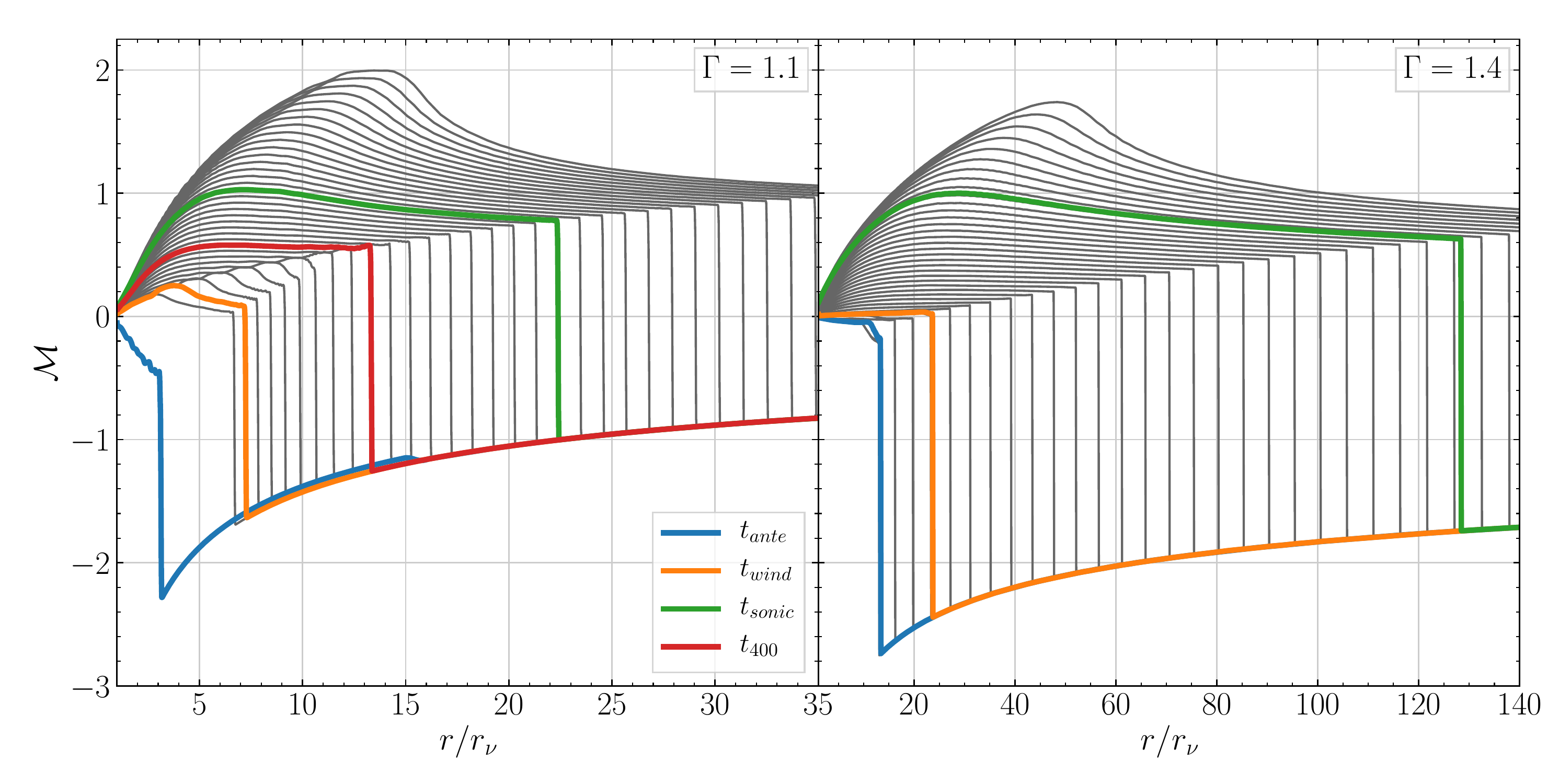}
                \caption{Mach number profiles for $\dot{M}_{\mathrm{acc},0}=1.06\:$M$_{\odot}\:$s$^{-1}$, with $\Gamma=1.1$ (left) and $\Gamma=1.4$ (right) cases, with the profiles corresponding to $t_{\mathrm{wind}},\;t_{\mathrm{ante}},$ and $t_{400}$ highlighted.  Output profiles are separated in time by 1 ms in the $\Gamma=1.1$ case, and 20 ms in the $\Gamma=1.4$ case.  The profile at $t=t_{400}$ is not shown for $\Gamma=1.4$, as the initial shock radius is near 400 km.  For $\Gamma=1.1$, see \autoref{fig:LateProfiles} and \S\ref{sec:TTW}, \S\ref{sec:WDS} for further evolution of the wind.}
                \label{fig:ExplosionTimesProfiles}
            \end{figure*}

            We plot the velocity profiles of the simulation at these times in \autoref{fig:ExplosionTimesProfiles}.  Though the $\Gamma=1.4$ case is larger in physical extent than the $\Gamma=1.1$ case, the velocity profiles are otherwise similar.  The $t_{400}$ profile is not shown for the $\Gamma=1.4$ case, as it coincidentally overlaps with the $t_{\mathrm{ante}}$ profile.

\begin{figure}
                \centering{}
                \includegraphics[width=\linewidth]{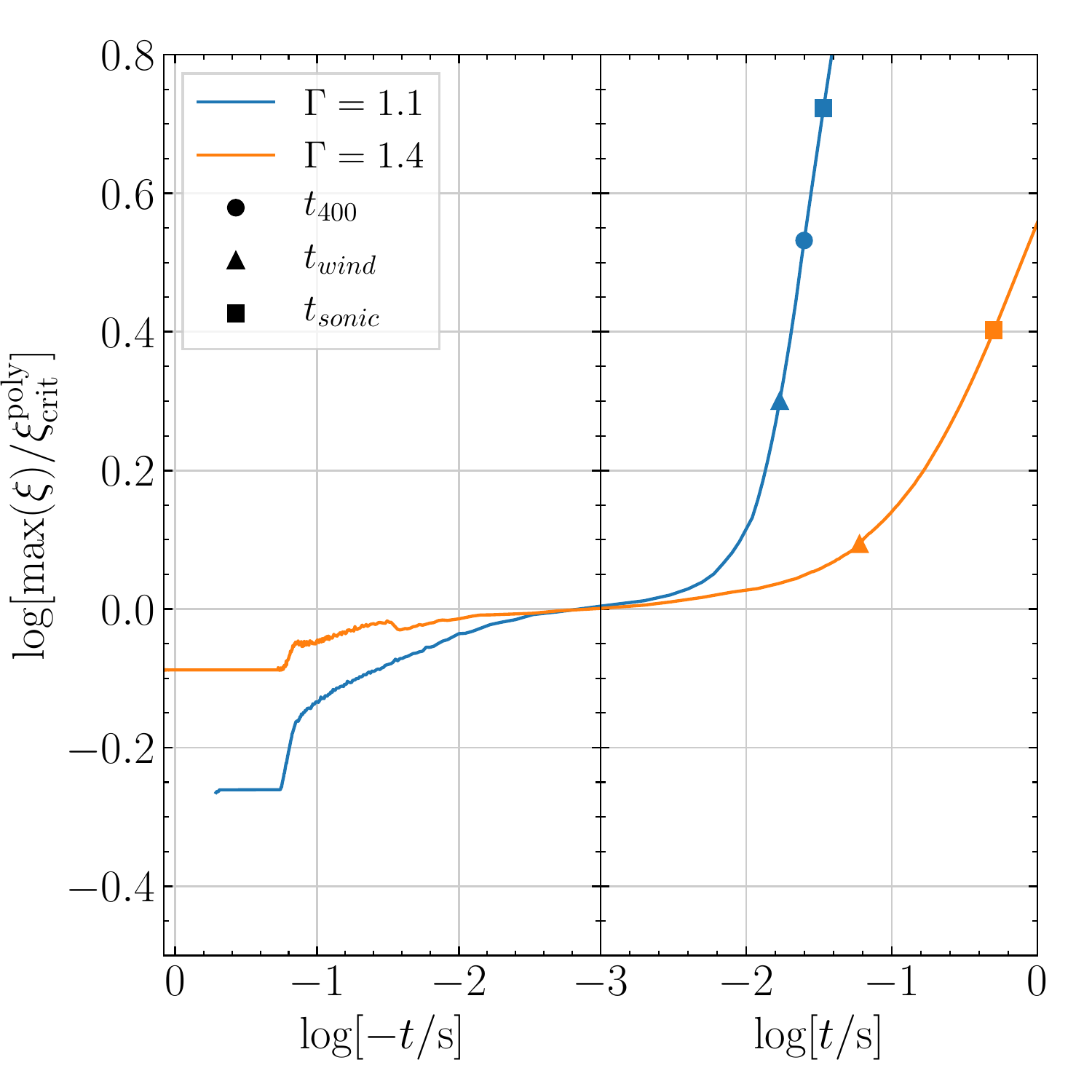}
                \caption{The maximum antesonic ratio vs. time for a configuration with $\dot{M}_{\mathrm{acc},0}=1.06\:$M$_{\odot}\:$s$^{-1}$, for both $\Gamma=1.1$ and $\Gamma=1.4$.  For comparison, the time-scales $t_{400}$, $t_{\mathrm{wind}}$, and $t_{\mathrm{sonic}}$ are also shown for each configuration.  The time-scale is normalized such that $t_{\mathrm{ante}}=0$; the left panel shows $t<0$, while the right panel shows $t>0$.  Both curves start at the point where the mass accretion rate is first reduced -- the period of constant $\xi$ at the beginning of the curves marks the time before this change in $\dot{M}_{\mathrm{acc}}$ has reached the shock. The growth of the antesonic ratio starts out slow in both cases, but quickly increases after the antesonic condition is met.  The $\Gamma=1.1$ simulation shows both a faster increase in $\xi/\xi_{\mathrm{crit}}$ and a larger maximum $\xi/\xi_{\mathrm{crit}}$.}
                \label{fig:AntesonicTimeseries}
            \end{figure}

            Finally, we also plot the (maximum) value of the antesonic ratio as a function of time in \autoref{fig:AntesonicTimeseries}.\footnote{The slight decrease in $\max(\xi)$ at $t\sim -0.03$  is due to the shock colliding with the trailing end of the $f_{\dot{M}}$ perturbation, which spreads out over many radial zones, and is steeper than the leading end. The sharp change in $\dot{M}$ manifests as a sharp change in $\rho$, which is visible in $\xi$ as $\xi\propto\rho^{\Gamma-1}$.}  We include markers at the three time-scales ($t_{400}$, $t_{\mathrm{wind}}$, and $t_{\mathrm{sonic}}$) identified above.  Not only does the explosion evolve much more slowly for $\Gamma=1.4$ than $\Gamma=1.1$, it does not reach as large of an antesonic ratio as the $\Gamma=1.1$ explosion does.

            Figures~\ref{fig:ExplosionTimesProfiles}~and~\ref{fig:AntesonicTimeseries} together indicate that rapid expansion of the shock coincides with rapid growth of the antesonic ratio, and that this phase of rapid evolution begins at $t_{\mathrm{ante}}$, or, at least, before $t_{\mathrm{wind}}$ or $t_{400}$.  Thus, we choose $t=t_{\mathrm{ante}}$ as a marker for the onset of explosion.

        \subsection{Resolution Dependence}\label{sec:Resolution}
            Here we investigate the effect of increasing resolution on  our ability to resolve the shock radius, the antesonic ratio, and the critical curve.  We measure the resolution of the study with the \emph{minimum} grid spacing, $\delta x$, which can be decreased by either changing the linear or adaptive refinement:
            \begin{equation}
                \delta x = \frac{x_{\mathrm{max}} - x_{\mathrm{min}}}{b\times n\times2^{\ell-1}} = \frac{621.25\:\mathrm{km}}{n\times2^{\ell}},
            \end{equation}
            where $n$ is the level of linear refinement, $b=16$ is the block size, and $\ell$ is the maximum number of AMR refinement levels.

\begin{figure*}
                \centering{}
                \includegraphics[width=\textwidth]{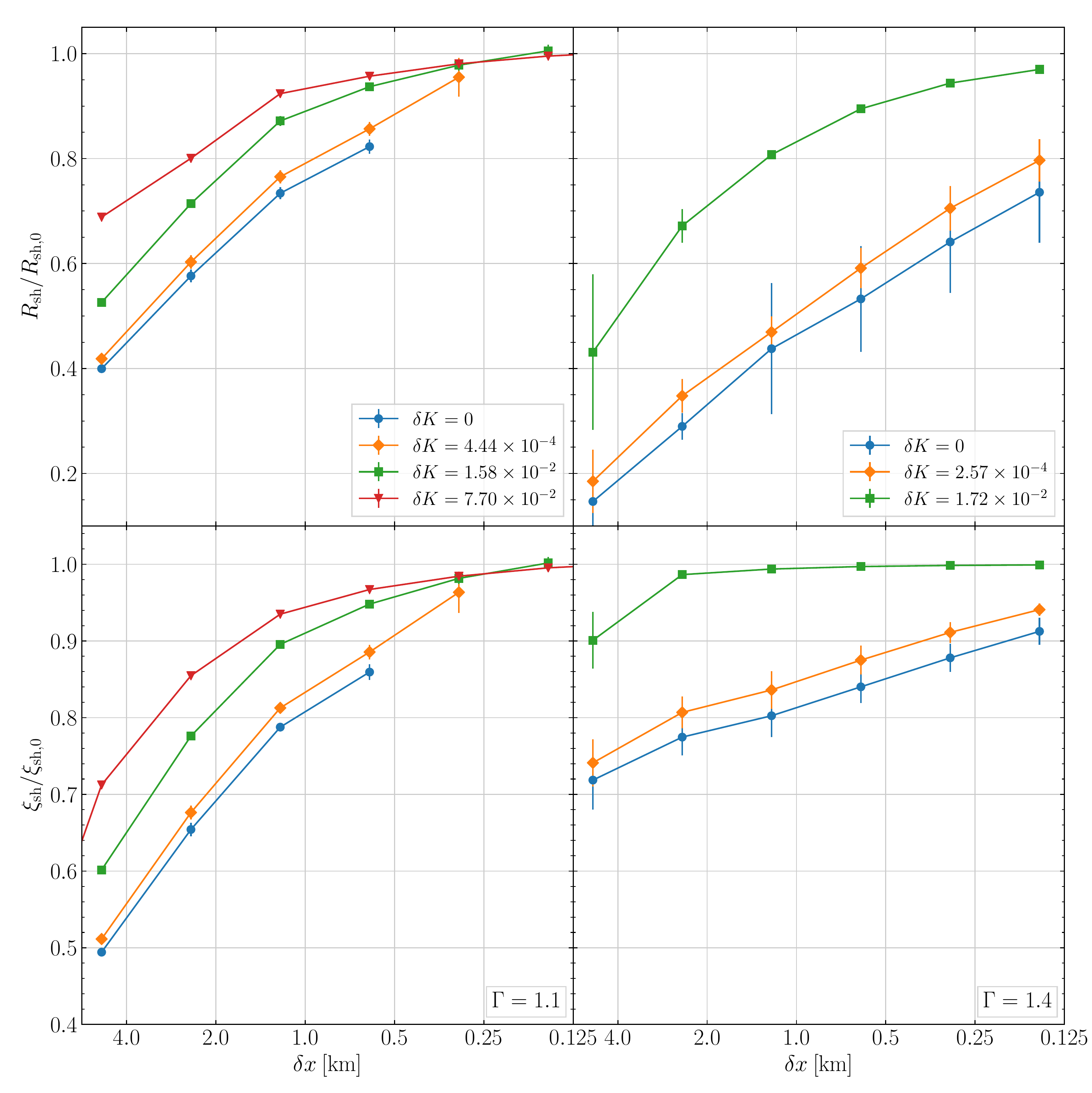}
                \caption{Shock radius (top) and antesonic ratio at the shock (bottom), relative to the PT12 value, for stable simulations at various distances from the critical curve, for $\Gamma=1.1$ (left) and $\Gamma=1.4$ (right), and with fixed $\dot{M} = 0.37\:$M$_{\odot}$.  The points show the mean value of the shock radius and antesonic ratio over the whole simulation time; error bars show the error on the mean.  We see that simulations close to the critical curve require higher resolution to converge to the same accuracy compared to those further away. Furthermore, $\Gamma=1.4$ simulations require higher resolution to achieve a similar accuracy in $R_{\mathrm{sh}}$ to $\Gamma=1.1$ simulations.  High resolution simulations for the smallest $\delta K$ cases exploded and are not plotted here; see the main text for our explanation of this behavior.}
                \label{fig:AntesonicConvergence}
            \end{figure*}

            In \autoref{fig:AntesonicConvergence}, we show the effect of increasing resolution (decreasing $\delta x$) on the fidelity of our simulations, as measured by the accuracy of the shock radius $R_{\mathrm{sh}}$ and the critical antesonic ratio relative to the value found by PT12.  We also show how this effect changes at various distances from the critical curve, defined as:
            \begin{equation}
                \delta K \equiv 1-\frac{K}{K_{c}}{}
                \label{eq:deltaK}
            \end{equation}
            where $K_{c}$ is the critical value of $K$ for the specified $\dot{M}_{\mathrm{acc}}$.  Thus, positive $\delta K$ (i.e., $K<K_{c}$) corresponds to solutions below the critical curve (i.e., steady accretion solutions), while negative $\delta K$  (i.e., $K>K_{c}$) corresponds to solutions above the critical curve (i.e., wind solutions).

            We find that under-resolved simulations systematically underpredict $R_{\mathrm{sh}}$ and $\xi_{\mathrm{sh}}$ and overestimate the critical curve, compared to the semi-analytic PT12 result.  Within the context of our simulations, this means that less well-resolved calculations will be more stable (less susceptible to explosion) than more highly resolved simulations.  Furthermore, simulations near the critical curve require higher resolution in order to calculate the shock radius to a given accuracy.  That is to say, at fixed physical resolution, configurations closer to the critical curve are less well resolved than configurations further away from the critical curve.

            \begin{figure*}
                \centering{}
                \includegraphics[width=\textwidth]{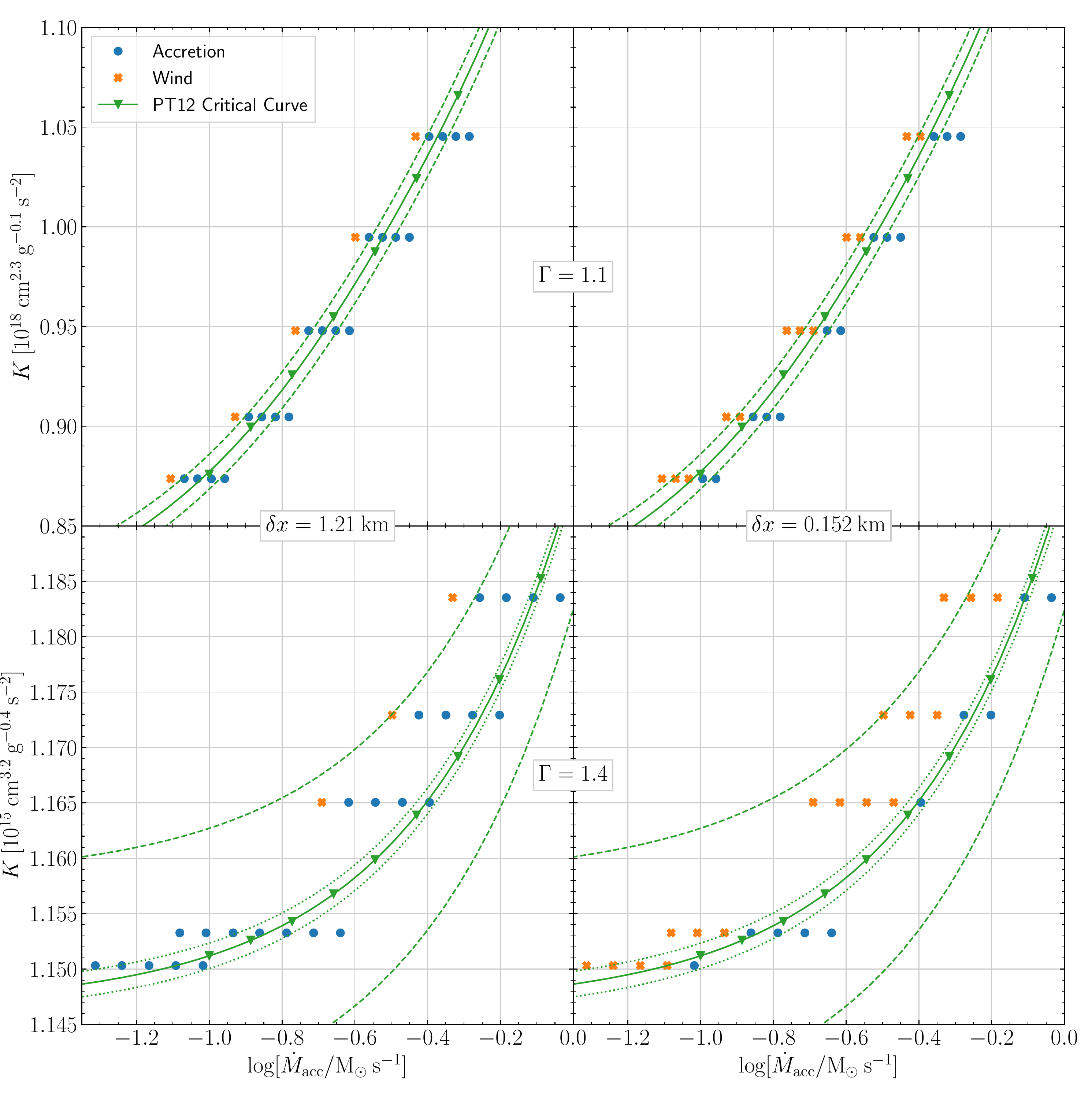}
                \caption{The critical curve at 2 different resolutions (left: below the fiducial resolution, right: above the fiducial resolution), for $\Gamma=1.1$ (top) and $\Gamma=1.4$ (bottom).  The simulations shown here are the same as in \autoref{fig:CriticalCurve} (there run at the fiducial resolution).  Green lines show curves of constant $\delta K=0$ (solid), $\delta K=\pm 10^{-2}$ (dashed), and $\delta K=\pm10^{-3}$ (dotted).  We see that underresolved simulations have a critical curve that has a larger normalization than the PT12 result; as resolution increases the critical curve normalization approaches the PT12 value.  This effect is more extreme in the $\Gamma=1.4$ case, with all of the simulations run at the fiducial resolution yielding accretion solutions at low resolution.}
                \label{fig:CCRes}
            \end{figure*}

            For example, a simulation on the critical curve $(\delta K \simeq 0)$ with resolution $\delta x = 0.607\:$km measures the antesonic ratio to about 15 per cent accuracy and the shock radius to about 20 per cent accuracy for $\Gamma=1.1$.  For $\Gamma=1.4$, $\delta x = 0.607\:$km yields an accuracy of about 45 per cent in $R_{\mathrm{sh}}$ and 15 per cent in $\xi_{\mathrm{sh}}$.  Whereas, a simulation of the same resolution but displaced from the critical curve by $\delta K=7.7\times10^{{-2}}$ measures the antesonic ratio to about 5 per cent accuracy and the shock radius to better than 1 per cent accuracy.

            Some high resolution simulations are missing from the $\Gamma=1.1$ panels in this figure for small $\delta K$.  These simulations exploded at these resolutions.  Though this is prima facie inconsistent with PT12, careful examination of the time-dependence of these calculations indicates that small time-dependent fluctuations in the solution (as is common with some Eulerian hydro schemes) instigates explosion in profiles very close to the critical curve.

            In \autoref{fig:CCRes}, we show the effect of resolution on the critical curve itself (compare to \autoref{fig:CriticalCurve}).  We see that, at low resolution, configurations can be above the PT12 critical curve but still yield accretion solutions, whereas at high resolution, these configurations yield wind solutions as expected.  This potentially explains why we see accretion solutions above the critical curve in \autoref{fig:CriticalCurve} for $\Gamma=1.4$.{}

            Taken together, our results indicate that changes in resolution have a small (few per-cent for $\Gamma=1.1$, few tenths of a per-cent for $\Gamma=1.4$) effect on the location of the critical curve itself, i.e., on the critical $K$ for a given $\dot{M}$ (\autoref{fig:CCRes}), but a larger effect (greater than ten per-cent) on our ability to accurately measure the properties of simulations at fixed $\delta K$ (\autoref{fig:AntesonicConvergence}).

            We perform a more limited version (specifically, $\Gamma=1.1$ only, and with fewer choices of $\delta x$) of this analysis for the HLL and HLLC hydro solvers, to ensure that our choice of solver does not unduly affect our results.  We find that these solvers yield statistically equivalent values of $R_{\mathrm{sh}}/R_{\mathrm{sh},0}$ and $\xi_{\mathrm{sh}}/\xi_{\mathrm{sh},0}$ at $\delta x\lesssim1.2\:\mathrm{km}$.  At $\delta x\gtrsim1.2\:\mathrm{km}$, the LLF solver yields higher accuracy in $R_{\mathrm{sh}}/R_{\mathrm{sh},0}$ and $\xi_{\mathrm{sh}}/\xi_{\mathrm{sh},0}$ by factors of approximately 1.5 and 2, respectively.  The critical curves produced with these solvers (i.e., in figures analogous to \autoref{fig:CCRes}) are identical at (a given $\delta x$) under all three solvers.  We also perform a limited ($\Gamma=1.1$ only) version of this analysis using uniform resolution rather than AMR; the analogue of \autoref{fig:AntesonicConvergence} made using these simulations is qualitatively identical to the one presented here.

            For context, we compare these results to the resolutions used in recent 3D simulations. \citep{Takiwaki2012} uses logarithmically spaced zones, with a fractional resolution of $\frac{\delta x}{x}\simeq0.02$ (this corresponds to a linear resolution $\delta x=2$ km at a radius of 100 km).  \citet{Lentz2015} and \citet{Fernandez2015} use smaller fractional resolutions, $\frac{\delta x}{x}\simeq0.014$ and $\frac{\delta x}{x}\simeq0.0045$ respectively (corresponding to linear resolutions 1.4 km and 0.45 km, respectively, at a radius of 100 km). \citet{Couch2013a} use a maximum resolution of 0.49 km, and \citet{Radice2016} perform a resolution study for resolutions up to 0.191 km (though they do not run a full simulation at this resolution due to the high computational cost).  While not an exhaustive list of the past decade of supernova simulations, these selected studies indicate that 3D supernova simulations are typically not run at resolutions fine enough for convergence of $R_{\mathrm{sh}}$ or max$(\xi_{\mathrm{crit}}^{\mathrm{poly}})$ to 10 per-cent near the critical curve.  We note that, while no specific resolution threshold we can provide is directly translatable to multi-D studies (especially those including turbulence, SASI, and convection), we see no reason that such models would converge significantly faster, or at significantly lower resolutions, than ours.

        \subsection{The Transonic Wind}\label{sec:TTW}
\begin{figure*}
                \centering{}
                \includegraphics[width=0.5\linewidth]{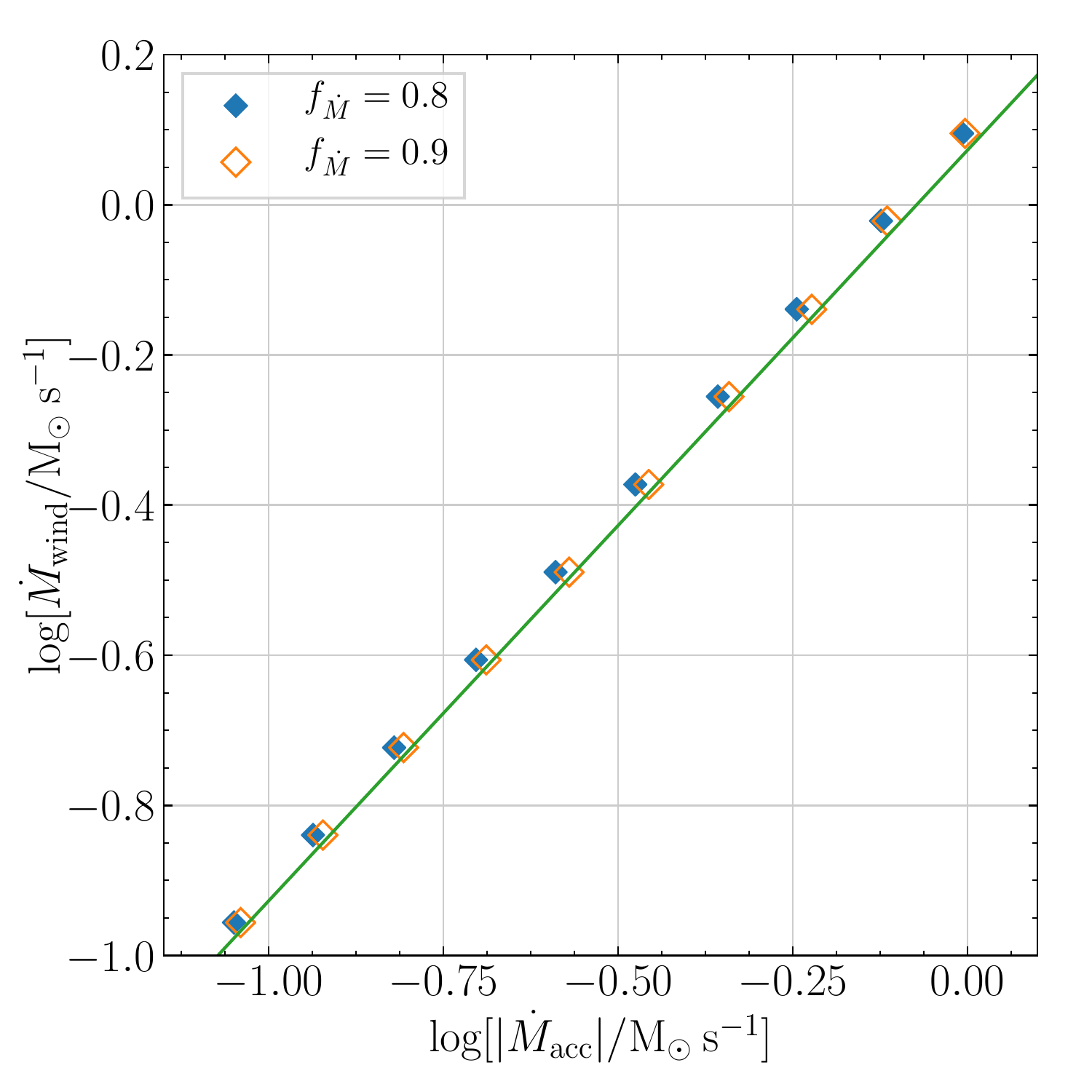}~
                \includegraphics[width=0.5\linewidth]{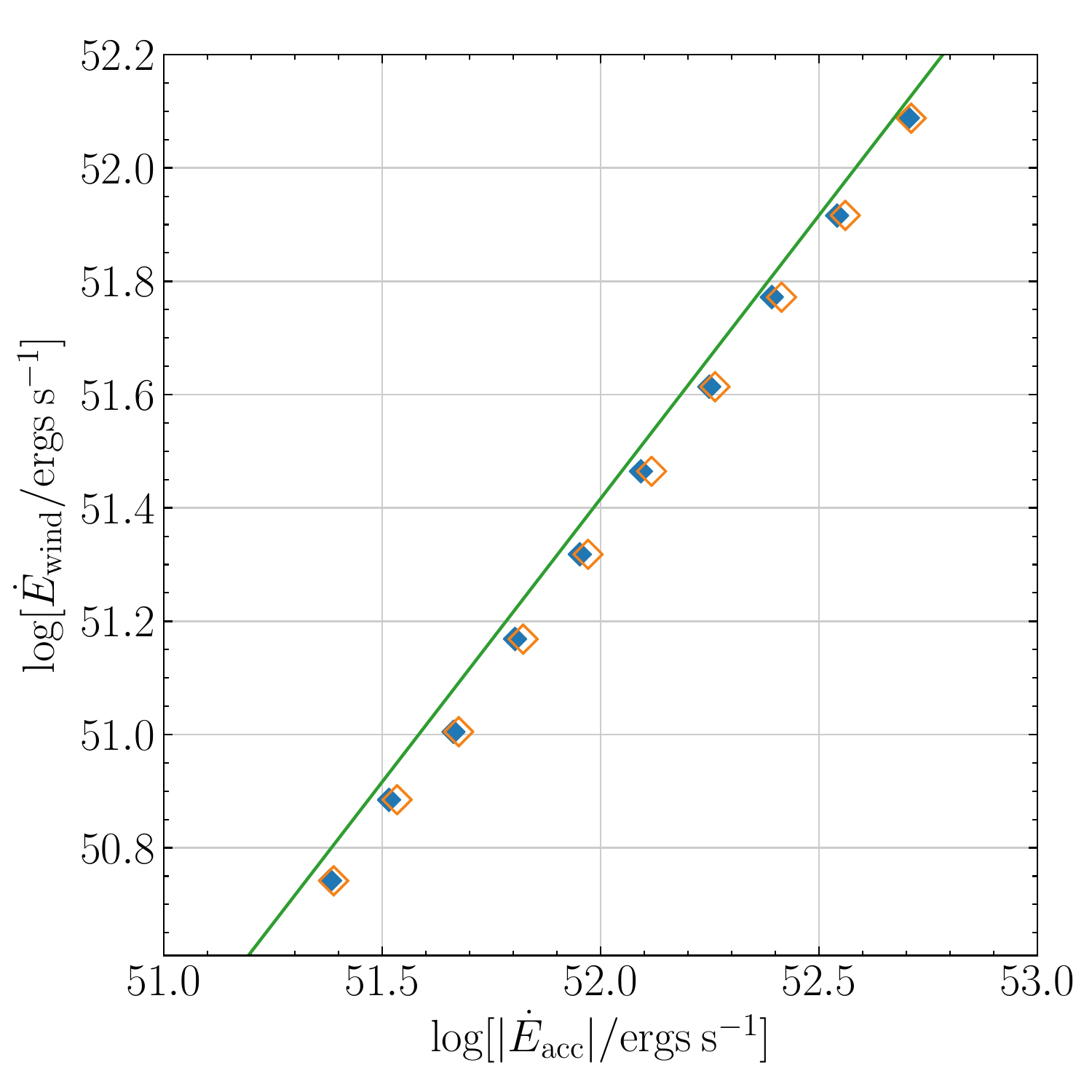}
                \caption{Left: the wind mass loss rate $\dot{M}_{\mathrm{wind}}$ of the post-explosion transonic wind as a function of the measured accretion rate $\dot{M}_{\mathrm{acc}}$, for $\Gamma=1.1$. Right: The kinetic power $\dot{E}=\frac{1}{2}\dot{M}v^{2}$ of the transonic wind, measured at the sonic point, versus that of the accretion flow, measured just in front of the shock at the onset of explosion.  The simulations here all have initial conditions on the critical curve, and have $f_{\dot{M}}=0.8$ and $0.9$, indicated by blue filled and orange unfilled diamonds respectively.  The green line is the analytic prediction obtained in \autoref{eq:mdotwa} and \autoref{eq:edotwa}, using the mean value of $\rho(R_\mathrm{sonic})/\rho(R_\mathrm{sh})$ across all simulations. We see that the wind mass loss rate and kinetic power do not depend on $f_{\dot{M}}$ -- each pair of simulations with the same $\dot{M}_{\mathrm{acc},0}$ but differing $f_{\dot{M}}$ has the same $\dot{M}_{\mathrm{wind}}$ and $\dot{E}_{\mathrm{wind}}$.  We also note that simulations with differing $f_{\dot{M}}$ have nearly the same measured $\dot{M}_{\mathrm{acc}}$ and $\dot{E}_{\mathrm{acc}}$.  Finally, we note that the simulations deviate slightly from our linear fit, implying that $\rho(R_\mathrm{sonic})/\rho(R_\mathrm{sh})$ has a small dependence on the accretion rate.  In actuality, the density ratio is a function of $K$, but $K$ and $\dot{M}_{\mathrm{acc}}$ are connected by the critical curve.}
                \label{fig:MdotWind}
            \end{figure*}

            Once the accretion flow exceeds the antesonic limit, it begins a time-dependent transition to a transonic wind, shown by the profiles in between the blue and green curves in \autoref{fig:ExplosionTimesProfiles}.  In \autoref{fig:MdotWind}, we plot the wind mass loss rate $\dot{M}_{\mathrm{wind}}$ against the mass accretion rate $\dot{M}_{\mathrm{acc}}$ at the onset of explosion (i.e., at $t=t_{\mathrm{ante}}$), and the kinetic power of the wind (as measured at the sonic point) $\dot{E}_{\mathrm{wind}}=\frac{1}{2}\dot{M}_{\mathrm{wind}}v(R_{\mathrm{sonic}})^{2}$ against the kinetic power of accretion (as measured just in front of the shock, at the onset of explosion) $\dot{E}_{\mathrm{acc}}=\frac{1}{2}\dot{M}_{\mathrm{acc}}v(R_{\mathrm{sh}})^{2}$.  We find that for configurations corresponding to initial conditions on the critical curve, $\dot{M}_{\mathrm{wind}}$ and $\dot{M}_{\mathrm{acc},0}$ are tightly correlated -- their ratio is very nearly constant. Furthermore, simulations with the same initial accretion rate $\dot{M}_{\mathrm{acc},0}$, but different decreases in $\dot{M}_{\mathrm{acc}}$, specified by $f_{\dot{M}}$, yield winds with the same $\dot{M}_{\mathrm{wind}}$, and have the same value of $\dot{M}_{\mathrm{acc}}$ at the onset of explosion.

            Naively, one might not expect such a tight correspondence between $\dot{M}_{\mathrm{wind}}$ and $\dot{M}_{\mathrm{acc}}$.  To see where this relationship comes from, note that the steady-state wind mass loss rate can be written as
            \begin{equation}
                \dot{M}_{\mathrm{wind}} = 4\pi R_{\mathrm{sonic}}^{2}c_{s}(R_\mathrm{sonic})\rho(R_{\mathrm{sonic}}) = \text{constant},\label{eq:dmwind_def}
            \end{equation}
            where
            \begin{equation}
                R_{\mathrm{sonic}} = \frac{GM}{2c_{s}^{2}(R_{\mathrm{sonic}})}\label{eq:rsonic}
            \end{equation}
            is the radius of the sonic point.  Using this relation, and the equation of state, we can write
            \begin{equation}
                \dot{M}_{\mathrm{wind}} = \frac{\pi G^{2}M^{2}}{(K\Gamma)^{3/2}}\rho(R_{\mathrm{sonic}})^{(5-3\Gamma)/2}.
            \end{equation}
            Similarly, we can write the mass accretion rate (which is constant everywhere) as
            \begin{equation}
                \dot{M}_{\mathrm{acc}} = 4\pi R_{\mathrm{sh}}^{2}v(R_{\mathrm{sh}})\rho(R_{\mathrm{sh}}),
            \end{equation}
            where we can use the critical condition to write the shock radius $R_{\mathrm{sh}}$ as
            \begin{equation}
                R_{\mathrm{sh}} = \frac{3GM\Gamma}{8c_{s}^{2}(R_{\mathrm{sh}})}.
                \label{eq:rshock_crit}
            \end{equation}
            Since the flow in front of the shock is in free-fall, the velocity just in front of the shock is given by
            \begin{equation}
                v(R_{\mathrm{sh}}) = -\sqrt{\frac{2GM}{R_{\mathrm{sh}}}} = -\frac{4}{\sqrt{3\Gamma}}c_{s}(R_{\mathrm{sh}}),
            \end{equation}
            we can write:
            \begin{equation}
                \dot{M}_{\mathrm{acc}} = -\frac{3\sqrt{3}\pi}{4}\frac{G^{2}M^{2}\Gamma^{3/2}}{(K\Gamma)^{3/2}}\rho(R_{\mathrm{sh}})^{(5-3\Gamma)/2}.
            \end{equation}
            Thus the ratio of the mass loss rate to the mass accretion rate is
            \begin{equation}
                \left|\frac{\dot{M}_{\mathrm{wind}}}{\dot{M}_{\mathrm{acc}}}\right| = \frac{4}{3\sqrt{3}}\frac{1}{\Gamma^{3/2}}\left(\frac{\rho(R_{\mathrm{sonic}})}{\rho(R_{\mathrm{sh}})}\right)^{(5-3\Gamma)/2}.
                \label{eq:mdotwa_ratio}
            \end{equation}
            For $\Gamma=1.1$, this is approximately equal to
            \begin{equation}
                \left|\frac{\dot{M}_{\mathrm{wind}}}{\dot{M}_{\mathrm{acc}}}\right| \simeq 0.667\left(\frac{\rho(R_{\mathrm{sonic}})}{\rho(R_{\mathrm{sh}})}\right)^{0.85},
                \label{eq:mdotwa}
            \end{equation}
            and, for $\Gamma=1.4$,
            \begin{equation}
                \left|\frac{\dot{M}_{\mathrm{wind}}}{\dot{M}_{\mathrm{acc}}}\right| \simeq 0.465\left(\frac{\rho(R_{\mathrm{sonic}})}{\rho(R_{\mathrm{sh}})}\right)^{0.4}.
                \label{eq:mdotwa_14}
            \end{equation}
            That is to say, we should expect the ratio of the wind mass loss rate to the accretion rate to be constant.  We see in \autoref{fig:MdotWind} that this is nearly true, with a small discrepancy due to the fact that $\rho(R_{\mathrm{sh}})$ and $\rho(R_{\mathrm{sonic}})$ have slight $K$ dependences.

            The implication of this result is that, when the antesonic condition is met, the kinetic power of the resulting wind is proportional to the accretion luminosity at the shock at the onset of explosion. Specifically, we expect:
            \begin{align}
                \left|\frac{\dot{E}_{\mathrm{wind}}}{\dot{E}_{\mathrm{acc}}}\right| &= \frac{v(R_{\mathrm{sonic}})^{2}}{v(R_{\mathrm{sh}})^{2}}\left|\frac{\dot{M}_{\mathrm{wind}}}{\dot{M}_{\mathrm{acc}}}\right|\\
                \left|\frac{\dot{E}_{\mathrm{wind}}}{\dot{E}_{\mathrm{acc}}}\right| &= \frac{1}{4\sqrt{3\Gamma}}\left(\frac{\rho(R_{\mathrm{sonic}})}{\rho(R_{\mathrm{sh}})}\right)^{(3-\Gamma)/2}.
            \end{align}
            That is, for $\Gamma=1.1$:
            \begin{equation}
                \left|\frac{\dot{E}_{\mathrm{wind}}}{\dot{E}_{\mathrm{acc}}}\right| \simeq  0.138\left(\frac{\rho(R_{\mathrm{sonic}})}{\rho(R_{\mathrm{sh}})}\right)^{0.95},\label{eq:edotwa}
            \end{equation}
            and, for $\Gamma=1.4$,
            \begin{equation}
                \left|\frac{\dot{E}_{\mathrm{wind}}}{\dot{E}_{\mathrm{acc}}}\right| \simeq  0.122\left(\frac{\rho(R_{\mathrm{sonic}})}{\rho(R_{\mathrm{sh}})}\right)^{0.8}.
            \end{equation}
            We see again that the ratio of $\dot{E}_{\mathrm{wind}}$ to $\dot{E}_{\mathrm{acc}}$ is very nearly constant, with a slight discrepancy due to the $K$ dependences of $\dot{E}_{\mathrm{wind}}$ and $\dot{E}_{\mathrm{acc}}$.  This discrepancy is naturally larger than in the wind mass loss/mass accretion case, as the exponent on $\rho(R_{\mathrm{sh}})/\rho(R_{\mathrm{sonic}})$ is larger in this case.

            In real supernovae, these relations only set the initial conditions for subsequent time evolution of the wind, which decreases in power as the PNS core cools. However, such cooling is not considered in this paper.

            The explanation for our observation that simulations with the same initial mass accretion rate, but differing $f_{\dot{M}}$, have the same measured accretion rate at the onset of explosion, is more subtle.  Though we initially specify a discontinuous jump in $\dot{M}_{\mathrm{acc}}$, this jump spreads out in radius as it propagates inwards, becoming a steep, but smooth, change in density when it encounters the shock.  Though the size and slope of this perturbation may change with $f_{\dot{M}}$, the flow will encounter the antesonic condition at the same $\dot{M}_{\mathrm{acc}}$ regardless of the choice of $f_{\dot{M}}$, because the simulations are using the same value of $K$, and thus the same critical $\dot{M}_{\mathrm{acc}}$.

            This is also true of our observation that $f_{\dot{M}}$ has no effect on $\dot{M}_{\mathrm{wind}}$.  In a real supernova, the collapsing star traces out a trajectory in the $(\dot{M}_{\mathrm{acc}},K)$ space (in reality, the $(\dot{M}_{\mathrm{acc}},L_{\nu})$ space) of \autoref{fig:CriticalCurve}, starting below the critical curve at high $\dot{M}_{\mathrm{acc}},$ $K$, and moving down as both $\dot{M}_{\mathrm{acc}}$ and $K$ decrease with time.  Shock revival occurs if and when the critical curve is crossed. Should the accretion flow hit a steep density jump, such as the density jumps between shell interfaces \citep{Pejcha2015,Summa2016,Ott2017} (or the smoothed $f_{\dot{M}}$ jump in our simulation), then the slope of this trajectory will become more shallow, leading to an intersection (and thus explosion) at higher $\dot{M}_{\mathrm{acc}}$ and $K$.  Conversely, should the accretion flow fail to hit such a density jump, then the explosion, if it occurs at all, will happen at lower $\dot{M}_{\mathrm{acc}}$, $K$.  In our simulations, then, since we keep constant $K$, the effect of $f_{\dot{M}}$ on the evolution of the accretion flow does not affect the mass accretion rate at which the flow reaches criticality.

            We stress that $K$, not $\dot{M}_{\mathrm{acc}}$, is the important factor in determining the evolution of the transonic wind.  The tight $\dot{M}_{\mathrm{wind}}$ -- $\dot{M}_{\mathrm{acc}}$ correlation is a consequence of the critical curve - the existence of the critical curve enforces a direct correspondence between $\dot{M}_{\mathrm{acc}}$ and $K$ at the onset of explosion.  Because our simulations do not have time-varying $K$, configurations of the same $\dot{M}_{\mathrm{acc},0}$ but different $f_{\dot{M}}$ explode with the same $K$, and thus naturally have the same transonic wind properties.

            This point is made clear when we look at initial conditions that lie significantly below the critical curve (i.e., configurations with the same $\dot{M}_{\mathrm{acc},0}$, but a smaller $K$, that are driven to explosion with larger perturbations, i.e., smaller $f_{\dot{M}}$.). These configurations are displaced from the $\dot{M}_{\mathrm{wind}}$ -- $\dot{M}_{\mathrm{acc}}$ correlation of the higher $K$ configurations - they have significantly smaller wind mass loss rates than the higher $K$ configurations of the same $\dot{M}_{\mathrm{acc},0}$ do.  However, when we investigate how the wind mass loss rate depends on $K$, we find that both the high $K$ and low $K$ configurations lie along the same relation.  That is to say, the properties of the wind are determined by the thermal properties of the model (i.e., the EOS parameters $K$ and $\Gamma$) rather than the accretion rate as specified by $\dot{M}_{\mathrm{acc}}$.

        \begin{figure*}
                \centering{}
                \includegraphics[width=\linewidth]{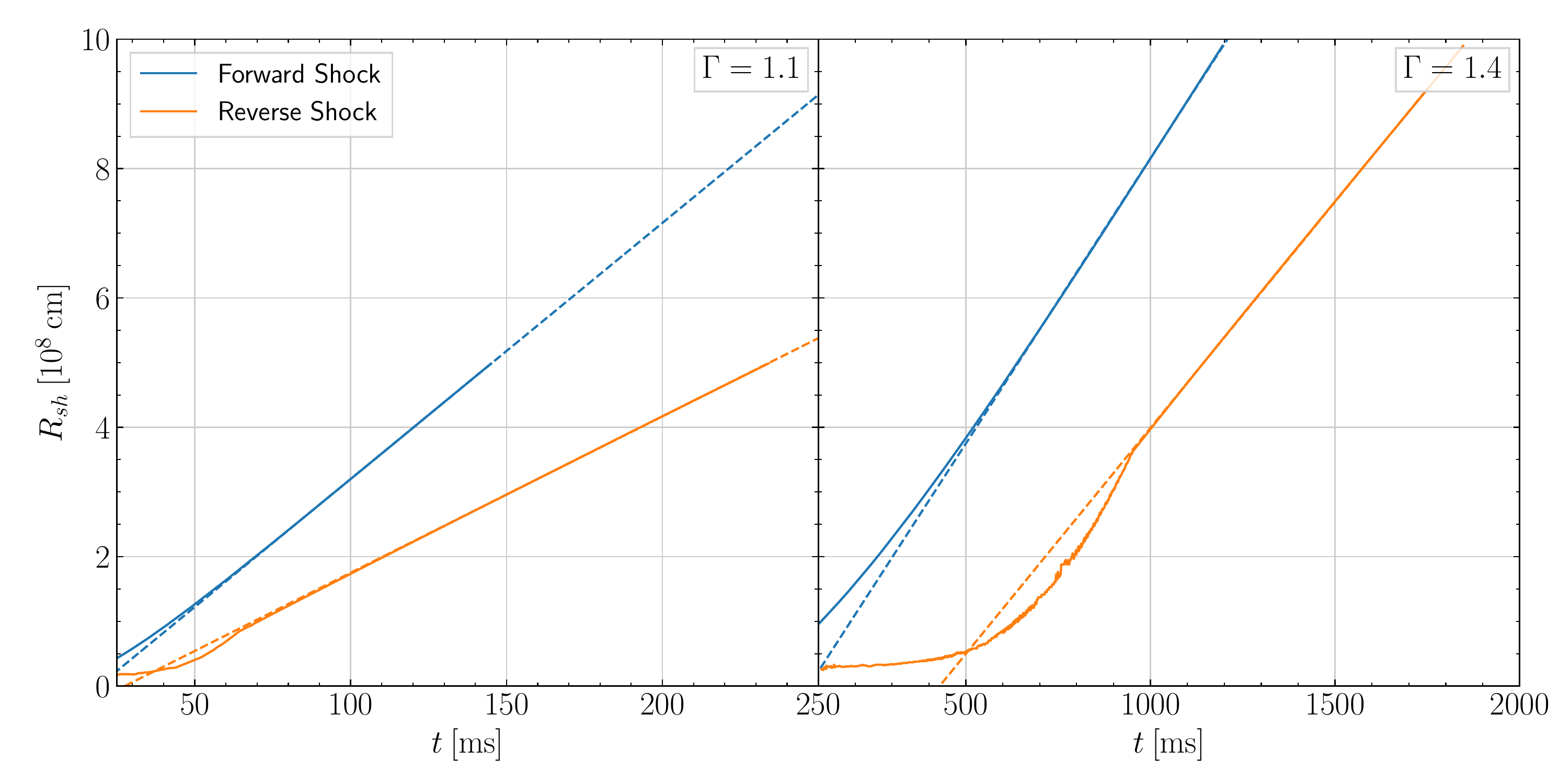}
                \caption{The radii of the forward and reverse shocks as a function of time, for $\Gamma=1.1$ (left) and $\Gamma=1.4$ (right), for $\dot{M}_{\mathrm{acc},0}=1.06\:$M$_{\odot}\:$s$^{-1}$. Dashed lines show linear fits to the later $t$ points; we see that the shock expansion is linear at these times -- i.e., that the shock velocity is constant.}
                \label{fig:PlugRadius}
            \end{figure*}

        \subsection{The Wind-Driven Shell}\label{sec:WDS}

            \begin{figure}
                \centering{}
                \includegraphics[width=\linewidth]{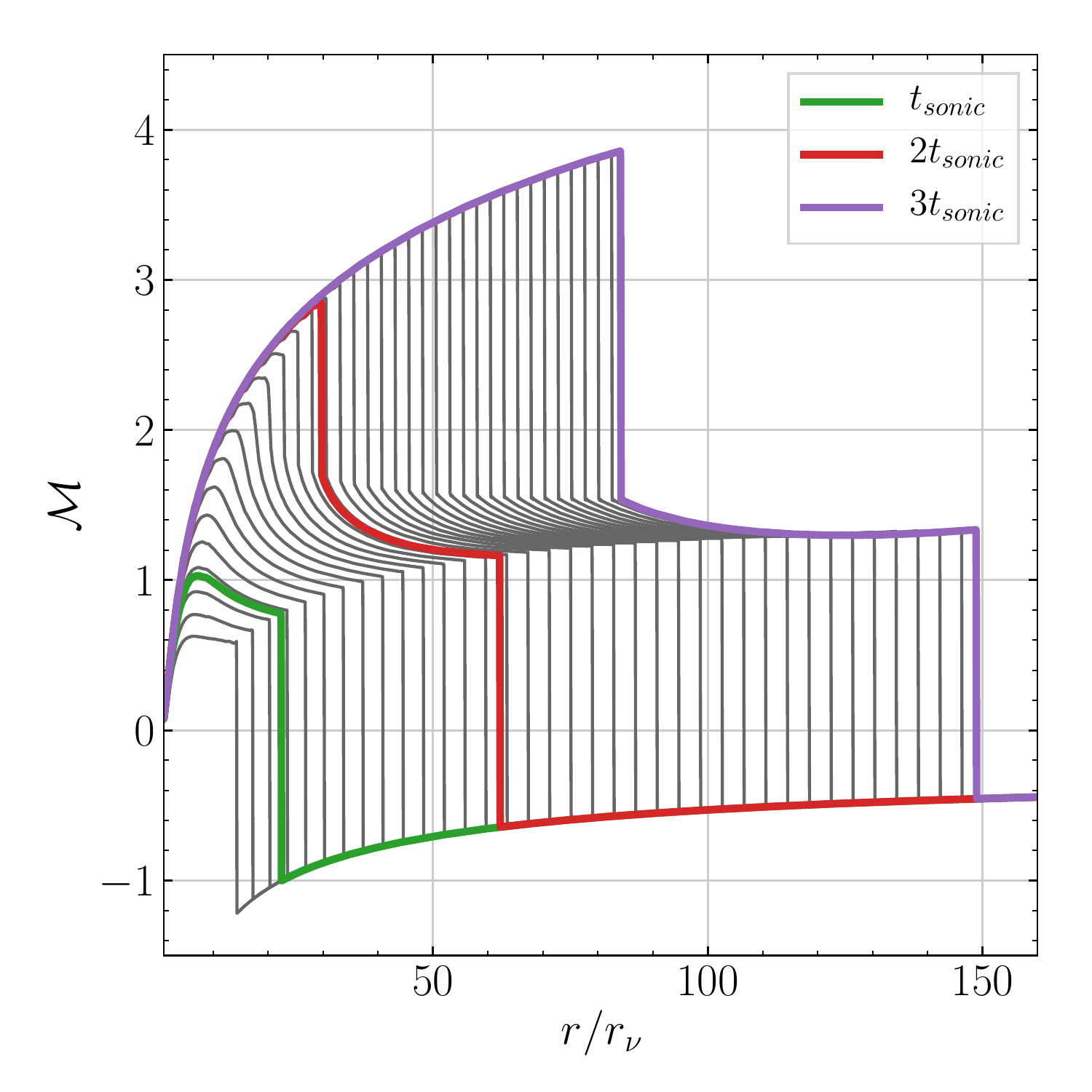}
                \caption{Mach number profiles for the $\dot{M}_{\mathrm{acc},0}=1.06\:$M$_{\odot}\:$s$^{-1}$ and $\Gamma=1.1$ model, as per \autoref{fig:ExplosionTimesProfiles}, but showing $t\gtrsim t_{\mathrm{sonic}}$ instead.  As before, the profile highlighted in green corresponds to $t=t_{\mathrm{sonic}}$.}
                \label{fig:LateProfiles}
            \end{figure}

            As the wind moves outward, it sweeps up the accreting gas, forming a high density peak significantly downstream of the forward shock, as seen in \autoref{fig:LateProfiles}.  This peak steepens with time and eventually forms a secondary, reverse shock.  We refer to the region in between the two shocks as the wind-driven shell.

            \autoref{fig:PlugRadius} shows the radius of the forward and reverse shocks as a function of time.  We see that the shocks quickly approach a constant velocity, and that the reverse shock moves more slowly than the forward shock -- the wind-driven shell grows larger over time.  This behavior is qualitatively similar to wind emergence seen in supernova models such as \citet{Burrows1995,Janka1996}.

        \subsection{Time-Dependent Perturbations}\label{sec:TDP}
\begin{figure}
                \centering{}
                \includegraphics[width=\linewidth]{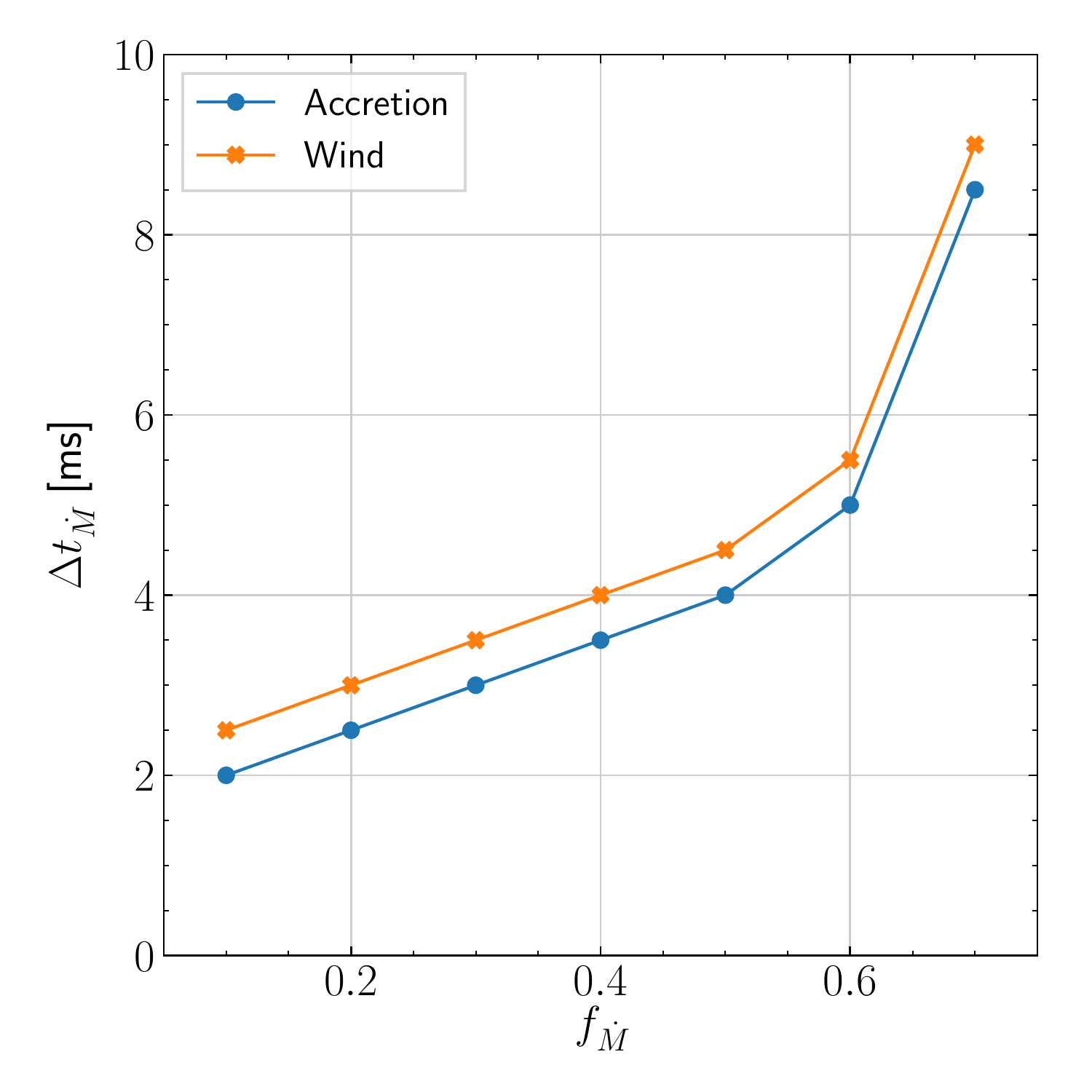}
                \caption{Stability of the simulation against transient perturbations, for an initial accretion rate $\dot{M}_{\mathrm{acc},0}=1.06\:\mathrm{M_{\odot}\:s^{-1}}$.  The initial conditions correspond to a configuration on the critical curve.  We see that for larger  decreases in accretion rate (i.e., for smaller $f_{\dot M}$), the maximum stable perturbation time is smaller.}
                \label{fig:TDPerturbations}
            \end{figure}

            Here we investigate the stability of various configurations to time-dependent perturbations of $\dot{M}_{\mathrm{acc}}$. We take initially stable configurations near, but below the critical curve, and reduce the mass accretion rate by a factor $f_{\dot{M}}$ for a short time $\Delta t_{\dot{M}}$.  We find that there is a threshold for the duration of this perturbation, below which, no explosion is produced despite the fact that the configuration is unstable for finite time.  Our results are summarized in \autoref{fig:TDPerturbations}. We see that the critical $\Delta t_{\dot{M}}$ changes quickly towards larger $f_{\dot{M}}$ (smaller perturbations), and changes more slowly towards smaller $f_{\dot{M}}$ (larger perturbations).

            These perturbations can be viewed as analogous to perturbations caused by asphericities in the Si/O burning layers found in 2D and 3D progenitor models \citep{Arnett2011,Couch2013a,Couch2015}.  These papers suggest characteristic perturbation magnitudes of $20$ per cent in velocity, corresponding to an equal size perturbation in $\dot{M}$ in our model (i.e., $f_{\dot{M}}=0.8$).  The exact magnitude of these perturbations should not be overly stressed; \citet{Mueller2015,Mueller2016} show that these perturbations are multi-dimensional and highly progenitor-dependent.  Furthermore, pre-collapse perturbations are not expected in all progenitors. As such, their overall impact on the general supernova problem is still unclear.

            These results can also be interpreted in the context of pre-shock turbulence.  Our results suggest that the perturbation to $\dot{M}_{\mathrm{acc}}$ required for explosion is dependent on the scale of the perturbations.  Larger-scale perturbations (lasting for a longer length of time) can lead to explosion for smaller total density perturbation, whereas smaller-scale perturbations (lasting for a shorter length of time) require a larger decrease in density.

            We find no evidence for oscillatory solutions, whether stable (i.e., remaining constant in amplitude) or unstable (i.e., increasing in amplitude until the critical curve is crossed), such as those found in \citet{Gabay2015} and \citet{Fernandez2012}.  However, as our polytropic EOS does not have any radially dependent heating or cooling terms, our lack of evidence is not in conflict with the \citet{Gabay2015} or \citet{Fernandez2012} results.

    \section{Conclusion}\label{sec:Conclusion}

        We perform simulations of simple supernova models with a polytropic equation of state in an effort to extend the findings of \citet{Pejcha2012} to time-dependent models.  Our findings are as follows:

        \begin{enumerate}
            \item{}
            We extend the analytic results of PT12 to derive a critical condition for a polytropic EOS (\autoref{eq:antepoly}, \autoref{fig:Euler}, and Appendix~\ref{app:antesonic}).
            \item{}
            We verify the existence of the antesonic condition in time-dependent simulations, and that it is the same as the time-steady antesonic condition (\S\ref{sec:TAC}; \autoref{fig:CriticalCurve} and \autoref{fig:CCRes}).  In particular, when the antesonic condition is exceeded, we observe a time-dependent evolution from accretion to a thermally driven wind (\S\ref{sec:TTW}; \S\ref{sec:WDS}; \autoref{fig:ExplosionTimesProfiles}; \autoref{fig:LateProfiles}).
            \item{}
            The value of the adiabatic index $\Gamma$ has a strong effect on the length and time scales of the evolution of the explosion.  Simulations using $\Gamma=1.4$ have significantly larger critical shock radii than simulations with $\Gamma=1.1$ do, and evolve more slowly (\S\ref{sec:TAC}; \autoref{fig:AntesonicTimeseries}).
            \item{}
            High resolution is required to fully capture the nature of the critical condition for explosion.  At low resolutions, the critical curve is shifted to higher $K$. Configurations that explode at high resolution fail to do so at lower resolution.  Only the highest resolutions found in the literature would yield accuracy $\lesssim10$ per cent in the antesonic ratio or shock radius at the critical curve, under our model (\S\ref{sec:Resolution}; \autoref{fig:AntesonicConvergence} and \autoref{fig:CCRes}).
            \item{}
            There is an important physical connection between the post-explosion wind and the pre-explosion accretion flow.  In particular, the mass loss rate of the transonic wind is (nearly) proportional to the initial accretion rate of the PNS, and the kinetic power of the wind is (nearly) proportional to the kinetic power of accretion immediately before explosion (\autoref{fig:MdotWind}).  This relationship is a consequence of the relationship between the mass accretion rate and the EOS parameter $K$ (itself directly related to the post-shock entropy) imposed by the critical curve -- the wind properties are set by $K$, the value of which implies a certain $\dot{M}_{\mathrm{acc}}$ due to the constraint of the critical condition.  This result implies that a higher accretion rate at the onset of explosion leads to an explosion with a larger wind mass loss rate and thus more kinetic power in the wind (\S\ref{sec:TTW}).
            \item{}
            Our model supernovae are sensitive to time-dependent perturbations.  We find that temporary decreases in the mass accretion rate can lead to explosion, even when the length of the perturbation is much smaller than the time required for the supernova to transition to the wind phase (\S\ref{sec:TDP}; \autoref{fig:TDPerturbations}).
        \end{enumerate}

        The analysis of the antesonic condition performed in this paper is fundamentally limited to 1D.  While we can compare to 2D and 3D results (as in \S\ref{sec:Resolution}), any comparisons made therein remain speculative.  Ultimately, we lack a full theory of the antesonic condition in multiple dimensions and with more realistic physics.  How the antesonic condition might respond to turbulence driven by SASI or convection, for example, is unknown.  Our simulations merely provide a hint of the effect of resolution on the ability to perform accurate simulations of  supernovae (\S\ref{sec:Resolution}), and of the relationship between the initial progenitor properties and the properties of the explosion and remnant (\S\ref{sec:TTW}).  Both of these are of relevance to full physics, multi-D supernova modeling.  However, future work explicitly investigating the antesonic condition with a more realistic physical setup is needed.

    \section{Acknowledgements}

        This research made use of the yt-project, a toolkit for analyzing and visualizing quantitative data \citep{Turk2011}. This research made use of matplotlib, a Python library for publication quality graphics \citep{Hunter2007}.

        The research of OP is currently supported by project PRIMUS/SCI/17 from Charles University.

    \bibliography{antesonic.bib}

    \appendix{}

    \section{The Polytropic Antesonic Condition}\label{app:antesonic}

        An analytic expression for the antesonic condition can only be found for an isothermal equation of state.  Even with a simple, polytropic equation of state, the equations become intractable.

        However, consider a `graphical' solution to the problem.  When the velocity accretion profiles are plotted in the $\xi$-$\mathcal{M}$ space (such as in \autoref{fig:Euler}), they must intersect the curve $\mathcal{M}_{-}$ (defined in \autoref{eq:machminus}) if they have a shock.  It stands to reason then that there must be a `last' solution -- a solution where any increase in $K$ yields a solution that does not intersect $\mathcal{M}_{-}$ -- i.e., a solution that doesn't have a shock, and, according to PT12, represents a configuration that \emph{must} undergo a dynamical transition to a wind (i.e., a supernova explosion).  By definition, the $\mathcal{M}$ profile of this solution must be tangent to $\mathcal{M}_{-}$ at the point of intersection.  Thus, we can find the critical antesonic ratio by determining the antesonic ratio for which $\frac{\partial\mathcal{M}}{\partial\xi}=\frac{\partial\mathcal{M}^{-}}{\partial\xi}$.

        We can write $\frac{\partial\mathcal{M}}{\partial\xi}$ as:
        \begin{equation}
           \frac{\partial\mathcal{M}}{\partial \xi} = \frac{\partial\mathcal{M}}{\partial r}\frac{\partial r}{\partial \xi} = \frac{1}{c_{s}}\left(\frac{\partial v}{\partial r} - \mathcal{M}\frac{\partial c_{s}}{\partial r}\right)\frac{\partial r}{\partial \xi},
        \end{equation}
        where $\xi$ is the antesonic ratio:
        \begin{equation}
           \xi = \frac{c_{s}^{2}r}{2GM}.
        \end{equation}
        Thus we can write:
        \begin{align}
           \frac{\partial\xi}{\partial r} &= \frac{c_{s}^{2}r}{2GM}\left(\frac{1}{r} + \frac{2}{c_{s}}\frac{\partial c_{s}}{\partial r}\right){}\\
           \frac{\partial r}{\partial \xi} = \left(\frac{\partial \xi}{\partial r}\right)^{-1} &= \frac{1}{\xi}\left(\frac{1}{r} + \frac{2}{c_{s}}\frac{\partial c_{s}}{\partial r}\right)^{-1}\\
           \frac{\partial r}{\partial \xi} &= \frac{c_{s}r}{\xi}\left(c_{s} + 2r\frac{\partial c_{s}}{\partial r}\right)^{-1},
        \end{align}
        and thus:
        \begin{equation}
           \frac{\partial\mathcal{M}}{\partial \xi} = \frac{r}{\xi}\frac{\frac{\partial v}{\partial r} - \mathcal{M}\frac{\partial c_{s}}{\partial r}}{c_{s} + 2r\frac{\partial c_{s}}{\partial r}}\label{eq:d_m}.
        \end{equation}
        From \autoref{eq:machminus}, we can determine:
        \begin{equation}
           \frac{\partial\mathcal{M}_{-}}{\partial\xi} = \frac{1-\sqrt{\frac{\Gamma}{\Gamma-4\xi}}}{4\xi^{3/2}}.
        \end{equation}
        Thus, the critical condition can be written as:
        \begin{align}
           \frac{\partial\mathcal{M}}{\partial\xi} - \frac{\partial\mathcal{M}_{-}}{\partial\xi} &= 0,\\
           \frac{r}{\xi}\frac{\frac{\partial v}{\partial r} - \mathcal{M}\frac{\partial c_{s}}{\partial r}}{c_{s} + 2r\frac{\partial c_{s}}{\partial r}} - \frac{1-\sqrt{\frac{\Gamma}{\Gamma-4\xi}}}{4\xi^{3/2}} &= 0.\label{eq:crit_unsub}
        \end{align}
        For convenience, we rewrite the Euler equations (\autoref{eq:eulerc} and \autoref{eq:eulerm}) in the form:
        \begin{align}
            \frac{\partial \rho}{\partial r} &= -\frac{\rho}{2r}\frac{\xi^{-1} - 4\mathcal{M}^{2}}{1-\mathcal{M}^{2}},\\
            \frac{\partial v_{r}}{\partial r} &= \frac{v_{r}}{2r}\frac{\xi^{-1} - 4}{1-\mathcal{M}^{2}}\label{eq:d_v}.
        \end{align}
        Assuming a polytropic equation of state, the sound speed is
        \begin{equation}
            c_{s}^{2} = \frac{\partial P}{\partial\rho} = \frac{\partial}{\partial\rho}(K\rho^{\Gamma}) = K\Gamma\rho^{\Gamma-1},
        \end{equation}
        and thus we have:
        \begin{align}
            \frac{\partial c_{s}}{\partial r} &= \frac{\partial}{\partial r}\left(\sqrt{K\Gamma\rho^{\Gamma-1}}\right)\\
            \frac{\partial c_{s}}{\partial r} &= \frac{\Gamma-1}{2}\frac{c_{s}}{\rho}\frac{\partial\rho}{\partial r}\\
            \frac{\partial c_{s}}{\partial r} &= \frac{\Gamma-1}{4}\frac{c_{s}}{r}\frac{\xi^{-1}-4\mathcal{M}^{2}}{1-\mathcal{M}^{2}}\label{eq:d_cs}.
        \end{align}
        The critical condition (\autoref{eq:crit_unsub}) is thus exactly given by
        \begin{equation}
            \frac{\mathcal{M}_{-}}{\xi}
            \frac{\frac{\Gamma+1}{4\xi} - 2 - \mathcal{M}_{-}^{2}(\Gamma-1)}
            {\mathcal{M}_{-}^{2}(2\Gamma-3) + 1 - \frac{\Gamma-1}{2\xi}} -
            \frac{1-\sqrt{\frac{\Gamma}{\Gamma-4\xi}}}{4\xi^{3/2}} = 0\label{eq:crit}.
        \end{equation}
        We find the solution to the equation to be exactly
        \begin{equation}
            \xi_{\mathrm{crit}}^{\mathrm{poly}} = \frac{3}{16}\Gamma\label{eq:sol}.
        \end{equation}

\end{document}